\documentclass[twocolumn]{aastex701}
\usepackage{amsmath}
\usepackage{subfigure}
\usepackage{xcolor}

\begin{document}

\title{Forecasting the Constraint on the Hu-Sawicki $f(R)$ Modified Gravity in \\the CSST $3\times2$pt Photometric Survey}

\correspondingauthor{Yan Gong}
\email{Email: gongyan@bao.ac.cn}

\author{Jun-hui Yan}
\affiliation{National Astronomical Observatories, Chinese Academy of Sciences, Beijing 100101, China}
\affiliation{University of Chinese Academy of Sciences, Beijing 100049, China}
\email{yanjh@bao.ac.cn}

\author{Yan Gong*} 
\affiliation{National Astronomical Observatories, Chinese Academy of Sciences, Beijing 100101, China}
\affiliation{University of Chinese Academy of Sciences, Beijing 100049, China}
\affiliation{Science Center for China Space Station Telescope, National Astronomical Observatories, \\Chinese Academy of Sciences, 20A Datun Road, Beijing 100101, China}
\email{gongy@bao.ac.cn}

\author{Qi Xiong}
\affiliation{National Astronomical Observatories, Chinese Academy of Sciences, Beijing 100101, China}
\affiliation{University of Chinese Academy of Sciences, Beijing 100049, China}
\email{xiongqi@bao.ac.cn}

\author{Xuelei Chen}
\affiliation{National Astronomical Observatories, Chinese Academy of Sciences, Beijing 100101, China}
\affiliation{University of Chinese Academy of Sciences, Beijing 100049, China}
\affiliation{Department of Physics, College of Sciences, Northeastern University, Shenyang 110819, China}
\affiliation{Centre for High Energy Physics, Peking University, Beijing 100871, China}
\affiliation{State Key Laboratory of Radio Astronomy and Technology, China}
\email{xuelei@cosmology.bao.ac.cn}

\author{Qi Guo}
\affiliation{National Astronomical Observatories, Chinese Academy of Sciences, 20A Datun Road, Beijing 100101, China}
\affiliation{School of Astronomy and Space Sciences, University of Chinese Academy of Sciences(UCAS),\\Yuquan Road NO.19A Beijing 100049, China}
\email{guoqi@nao.cas.cn}

\author{Ming Li}
\affiliation{National Astronomical Observatories, Chinese Academy of Sciences, 20A Datun Road, Beijing 100101, China}
\email{mingli@nao.cas.cn}

\author{Yun Liu}
\affiliation{National Astronomical Observatories, Chinese Academy of Sciences, 20A Datun Road, Beijing 100101, China}
\affiliation{School of Astronomy and Space Sciences, University of Chinese Academy of Sciences(UCAS),\\Yuquan Road NO.19A Beijing 100049, China}
\email{liuyun@nao.cas.cn}

\author{Wenxiang Pei}
\affiliation{Shanghai Key Lab for Astrophysics, Shanghai Normal University, Shanghai 200234, China}
\email{wxpei@shnu.edu.cn}

\begin{abstract}
We forecast the constraint on the Hu-Sawicki $f(R)$ model from the photometric 
survey operated by the Chinese Space Station Survey Telescope (CSST). The 
 simulated $3\times2$pt data of galaxy clustering, weak lensing,
and galaxy-galaxy lensing measurements within 100 deg$^{2}$ are used in the analysis. 
The mock observational maps are constructed from a
light cone, redshift sampling and noise. The angular power
spectra are measured with pseudo-$C_\ell$ estimators and compared to theory in
the same basis using validated weighting functions and an analytic covariance matrix that
includes Gaussian, connected non-Gaussian, and super-sample terms. We model the 
theoretical spectra using two methods. The first one uses \textsc{MGCAMB} to
compute the linear modified-gravity clustering power spectra, and the second one adopts the \textsc{FREmu} emulator with a baseline of nonlinear $\Lambda$CDM prescription. Parameter inference is performed with \textsc{Cobaya}, and the cosmological and modified-gravity parameters are sampled within the emulator training domain, which are jointly fitted with the systematic parameters. We find that the predictions from the two methods are in good agreement at the overlapping large scales, and the emulator method can correctly provide additional high-$\ell$ information.
The $1\sigma$ upper bounds of $\log_{10}|f_{R0}|$ are found to be $<-5.42$ for cosmic shear only case and $<-5.29$ for the 100 deg$^2$ CSST $3\times2$pt probe. The full CSST photometric survey with 17,500 deg$^2$ survey area is expected to further improve the constraint precision by about one order of magnitude. Our results demonstrate that the CSST $3\times2$pt survey can deliver strict tests on $f(R)$ gravity.
\end{abstract}



\section{Introduction}\label{sec:intro}

The discovery of late-time cosmic acceleration
through observations of Type Ia supernovae (SNe Ia)
\citep{Perlmutter1999,Riess1998}
has profoundly challenged our understanding
of fundamental physics.
The prevailing explanation invokes dark energy 
a component with negative pressure
driving the accelerated expansion.
The simplest realization is the cosmological
constant $\Lambda$ together with
cold dark matter ($\Lambda$CDM)  
provides an excellent explanation to a broad suite
of cosmological observations,
including the anisotropies of cosmic microwave 
background (CMB) \citep[e.g.][]{Planck2018}, cosmic 
large-scale structure (LSS) \citep[e.g.][]{Alam2021},
and SN Ia distances \citep[e.g.][]{Scolnic2018}.

Nonetheless, the cosmological constant
raises significant theoretical puzzles,
such as the fine-tuning and coincidence problems.
Accordingly, another possibility is that the
observed acceleration reflects a departure
from Einstein's general relativity (GR)
on cosmological scales rather than the
presence of a new exotic component.
This leads to the idea of modified gravity 
theories, in which the law of gravity itself 
may differ at large scales, yielding 
accelerated expansion without invoking an 
additional dark energy fluid \citep{Clifton2012}.

Among modified gravity proposals,
$f(R)$ theories are one of the simplest
and most extensively studied extensions
\citep{Buchdahl1970,Starobinsky1980}.
In these models, the Einstein-Hilbert action
is generalized to include a function
of the Ricci scalar $R$.
As a concrete example, we consider the
Hu-Sawicki parameterization of $f(R)$ gravity
\citep{Hu2007}.
This class of models introduces an
additional scalar degree of freedom,
which alters the relation between matter
overdensities and metric potentials,
and thereby modifies the growth history
and power spectrum of large scale structure
relative to GR.

Consequently, probes of structure formation 
including galaxy clustering, weak lensing,
cluster abundances, and redshift-space
distortions, provide direct and
complementary tests of $f(R)$ modifications.
Current third-generation (Stage-III) surveys
have already produced representative bounds
on the Hu-Sawicki parameter, commonly
expressed as upper limits on
$\log_{10}|f_{R0}|$.
Analyses of cosmic shear from KiDS-1000
suggest that when $f_{R0}$ is included
in nonlinear modeling, the data alone
tend to yield weak, prior-dominated constraints
\citep{Troester2021}.
Similarly, HSC-Y1 cosmic-shear studies
which employ $k$-cut techniques to suppress
small-scale modeling uncertainties report
constraints that remain largely
prior dominated \citep{Vazsonyi2021}.

Studies that combine weak lensing with
other probes, e.g., cluster abundances
with weak-lensing mass calibration, have
achieved substantially stronger limits.
For example, a recent joint analysis using
SPT tSZE-selected clusters calibrated by
Dark Energy Survey (DES) and Hubble Space 
Telescope (HST) weak lensing reports
$\log_{10}|f_{R0}| < -5.32$ (95\% CL)
\citep{SPT:2024adw}.
More generally, combinations of Stage-III
weak lensing (DES-Y3, KiDS-1000, HSC-Y3)
with external datasets can push the
upper bound on $\log_{10}|f_{R0}|$
into the $-4$ to $-5$ range,
with the precise limit sensitive to the choices
in nonlinear modeling and screening implementation
\citep{Bai2024}.

In this work we focus on the 3$\times$2pt
 photometric observation of a
forthcoming Stage-IV survey, i.e. the
Chinese Space Station Survey Telescope (CSST)
\citep{Zhan2021,Gong:2019yxt, CSST:2025ssq, Gong:2025ecr}.
Our goal is to investigate the capability of constraining 
the Hu-Sawicki $f(R)$ parameter with the CSST
galaxy clustering and weak-lensing
measurements alone (i.e., without CMB
or other external probes),
and evaluate CSST's potential
to distinguish modified gravity
from the $\Lambda$CDM paradigm.

The structure of this paper is organized 
as follows. In Section~\ref{sec:theory}, we introduce 
the theoretical framework of the Hu-Sawicki $f(R)$ 
gravity model and derive the angular power spectrum 
formalism for the $3\times2$pt analysis, including 
the treatment of systematic effects. 
Section~\ref{sec:data} describes the construction 
of mock observational data from the CSST photometric 
survey, including galaxy clustering, cosmic shear, 
and galaxy-galaxy lensing measurements, along with 
the covariance matrix estimation. In 
Section~\ref{sec:constraints}, we present the parameter 
constraint methodology, and report the constraints 
on the Hu-Sawicki parameter $\log_{10}|f_{R0}|$ and 
other cosmological parameters. Finally, 
Section~\ref{sec:summary} summarizes our main findings 
and discusses the potential of the full CSST survey for 
testing modified gravity theories.

\section{Theoretical Framework}
\label{sec:theory}
\subsection{The Hu-Sawicki Model of $f(R)$ Gravity}
$f(R)$ theories extend the Einstein-Hilbert action
by replacing the Ricci scalar $R$ with a general
function $f(R)$, thereby introducing an extra
scalar degree of freedom that can mediate
a long-range fifth force and requires screening
mechanisms (e.g.\ the chameleon) to satisfy
local gravity tests \citep{Hu2007,Khoury2004,DeFelice2010}.
The total action of $f(R)$ gravity can be expressed as
\begin{equation}
S=\int d^{4}x\,\sqrt{-g}\,
\left[\frac{f(R)}{16\pi G}\right]+S_{\rm m}+S_{\rm r},
\label{eq:action}
\end{equation}
where $S_{\rm m}$ and $S_{\rm r}$ denote the
matter and radiation actions, respectively.
Variation of Eq.~\eqref{eq:action} with respect
to the metric yields the modified field equations:
\begin{equation}
f_{R}R_{\mu\nu}-\tfrac{1}{2}f\,g_{\mu\nu}
-\nabla_{\mu}\nabla_{\nu}f_{R}
+g_{\mu\nu}\,\Box f_{R}=8\pi G\,T_{\mu\nu},
\label{eq:field_eq}
\end{equation}
with $f_{R}\equiv df/dR$ and
$\Box\equiv g^{\alpha\beta}\nabla_{\alpha}\nabla_{\beta}$.

Assuming a spatially flat 
FLRW metric,
the Ricci scalar is $R=6\,(2H^{2}+\dot{H})$,
where $H\equiv\dot{a}/a$ and dots denote
derivatives with respect to cosmic time $t$.
The modified Friedmann equations become
\begin{align}
3f_{R}H^{2}
&=8\pi G\,(\rho_{\rm m}+\rho_{\rm r})
+\tfrac{1}{2}(f_{R}R-f)-3H\,\dot{f}_{R},
\label{eq:fried1} \\
-2f_{R}\,\dot{H}
&=8\pi G\,(\rho_{\rm m}+p_{\rm m}
+\rho_{\rm r}+p_{\rm r})
+\ddot{f}_{R}-H\,\dot{f}_{R}.
\label{eq:fried2}
\end{align}

In the following analysis, 
we neglect the contributions from 
radiation ($\rho_{\rm r}$ and $p_{\rm r}$) 
as we focus on the late-time universe.
It is convenient to convert the geometric modifications
into an effective energy density and pressure which are defined by
\begin{align}
\rho_{\rm eff}
&=\frac{1}{8\pi G}\Big[\tfrac{1}{2}(f_{R}R-f)
-3H\,\dot{f}_{R}+3(1-f_{R})H^{2}\Big],
\label{eq:rho_eff} \\
p_{\rm eff}
&=\frac{1}{8\pi G} \times \Big[-\tfrac{1}{2}(f_{R}R-f)+\ddot{f}_{R}+2H\,\dot{f}_{R}  \nonumber \\
& -(1-f_{R})(2\dot{H}+3H^{2})\Big].
\label{eq:p_eff}
\end{align}
With these definitions, the Friedmann equation can 
assume the familiar form
$3H^{2}=8\pi G(\rho_{\rm m}+\rho_{\rm eff})$,
and the effective equation of state is defined by
$w_{\rm eff}\equiv p_{\rm eff}/\rho_{\rm eff}$
\citep{Sotiriou2010,DeFelice2010}.

Under the quasi-static and sub-horizon approximation, 
i.e. neglecting time derivatives of perturbations
relative to spatial derivatives, the linearized
field equations in Newtonian gauge reduce to
a modified Poisson equation and a scalaron
(field) equation. Using the comoving Laplacian
$\nabla^{2}$, the following relations can be derived as
\citep{Hu2007,DeFelice2010,Khoury2004}
\begin{equation}
\nabla^{2}\Phi=\frac{16\pi G}{3}\,\delta\rho
-\frac{1}{6}\,\delta R,
\label{eq:poisson_fR}
\end{equation}
\begin{equation}
\nabla^{2}\,\delta f_{R}
=\frac{1}{3}\,\bigl(\delta R-8\pi G\,\delta\rho\bigr),
\label{eq:scalaron_eq}
\end{equation}
where $\delta f_{R}\equiv f_{RR}\,\delta R$ with
$f_{RR}\equiv d^{2}f/dR^{2}$, and
$\delta R$ and $\delta\rho$ denote perturbations
of the Ricci scalar and matter density.

\begin{figure}
    \centering
    \includegraphics[width=0.95\linewidth]{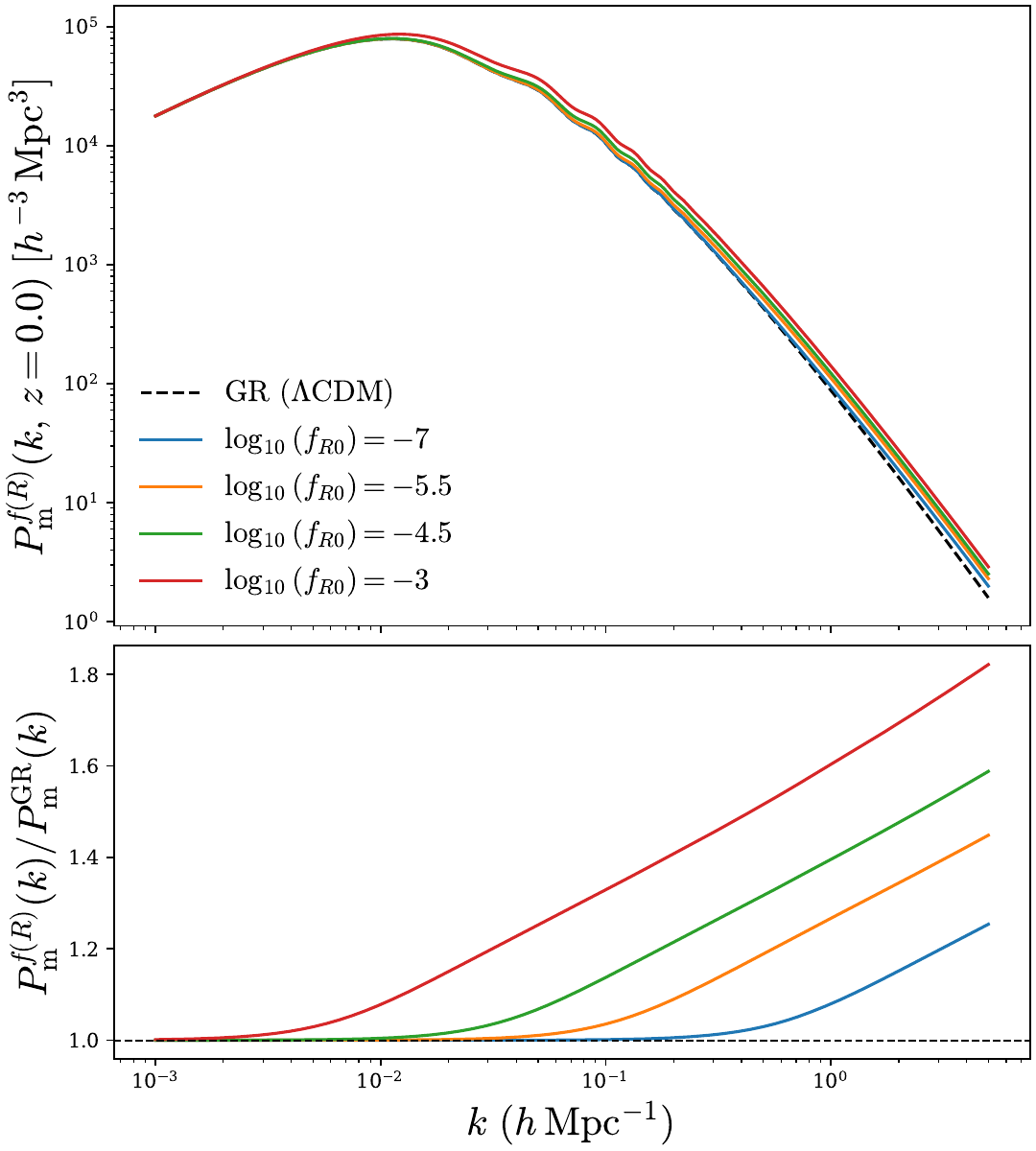}
    \caption{{\it Upper panel}: The linear matter power spectrum
    $P_{\rm m}(k)$ at $z=0$ for different values of 
     $\log_{10}|f_{R0}|$ and the $\Lambda$CDM model calculated by \textsc{MGCAMB}.
     {\it Lower panel}: The ratio $P^{f(R)}_{\rm m}(k)/P^{\rm \Lambda CDM}_{\rm m} (k)$ for different values of $\log_{10}|f_{R0}|$.
     }
    \label{fig:Pk_comparison}
\end{figure}

\cite{Hu2007} proposed an $f(R)$ model that employs 
the chameleon mechanism to recover GR locally
on  high-density region, which can be parameterized as
\begin{equation}
f(R)=R-m^{2}\frac{c_{1}(R/m^{2})^{n}}
{c_{2}(R/m^{2})^{n}+1},
\label{eq:hu_sawicki}
\end{equation}
with $m^{2}\equiv 8\pi G\,\bar{\rho}_{\mathrm {m}0}/3
=H_{0}^{2}\Omega_{\rm m0}$ and $c_{1}$, $c_{2}$ and $n$ 
are the dimensionless parameters.
For $R\gg m^{2}$ the expansion yields
\begin{equation}
f(R)\simeq R-\frac{c_{1}}{c_{2}}\,m^{2}
+\frac{c_{1}}{c_{2}^{2}}\,m^{2}\left(\frac{m^{2}}{R}\right)^{n},
\end{equation}
so the background evolution can mimic
$\Lambda$CDM with $2\Lambda\simeq(c_{1}/c_{2})m^{2}$
when the last term is negligible \citep{Hu2007}.

The scalar degree of freedom $f_{R}\equiv df/dR$
deviates slightly from unity in the large-$R$ limit,
and the deviation can be written as
\begin{equation}
	\tilde{f}_{R}
\simeq 1 - \,n\,\frac{c_{1}}{c_{2}^{2}}
\left(\frac{m^{2}}{R}\right)^{n+1}.
\label{eq:fR_approx}
\end{equation}
Here we fix $n=1$ to simplify the model and facilitate comparison with a wide range of existing studies.
The present-day amplitude of this
deviation can be defined by 
$f_{R0}\equiv\tilde{f}_{R}(z=0)$. 
Small $|f_{R0}|$ indicates close agreement with GR on cosmological scales. The influence of $\log_{10}|f_{R0}|$ on the matter power spectrum at $z=0$ is presented in Figure~\ref{fig:Pk_comparison}, which shows a key observational feature of the enhanced power at small scales. As $\log_{10}|f_{R0}|$ decreases, the chameleon screening mechanism becomes more efficient, suppressing this enhancement and causing the spectrum to approach the $\Lambda$CDM case. We can see that the deviation from the $\Lambda$CDM model becomes significant at $k\gtrsim0.002$ or $0.2$ $h\, \mathrm{Mpc}^{-1}$ for $\log_{10}|f_{R0}|=-3$ or $-7$.

\subsection{Angular Power Spectrum}
\label{subsec:systematics}

Accurate modeling of the observed angular power spectra  
requires careful treatment of several observational systematics  
that can bias cosmological inference.  
These include photometric redshift uncertainties, shear calibration,  
galaxy bias, and intrinsic alignments (IA) of galaxy shapes.  

Photometric redshift uncertainties are modeled using  
the shift ($\Delta_z^i$) and stretch ($\sigma_z^i$)  
parameters for each tomographic bin $i$,  
which modify the estimated redshift distributions as \citep{Xiong2025}
\begin{equation}
n^i(z) \rightarrow \frac{1}{\sigma_z^i} \,
n^i\!\left( \frac{z - \langle z^i \rangle - \Delta_z^i}{\sigma_z^i}
+ \langle z^i \rangle \right).
\end{equation}
Here, the superscript $i$ denotes the redshift bin index,  
and $\langle z^i \rangle \equiv \int z \, n^i(z) \, dz$  is the mean redshift of that bin.  
Shear calibration biases are included as the multiplicative factor 
$(1+m_i)$ for each tomographic bin. 
Intrinsic alignment effect arises from  
correlations with the large scale tidal field,  
introducing spurious signals in shear auto- and cross-correlations.  
We adopt the non-linear alignment (NLA) model 
\citep{Hildebrandt:2016iqg, Johnston_2019},  
in which the IA contribution for the $i$-th bin is given by
\begin{equation}
F^i(\chi) = -A_{\text{IA}} \, C_1 \, \rho_c
\frac{\Omega_{\rm m}}{D(\chi)} \,
n^i(z(\chi)) \frac{dz}{d\chi}
\left( \frac{1+z}{1+z_0} \right)^{\eta_{\text{IA}}},
\end{equation}
where $A_{\text{IA}}$ and $\eta_{\text{IA}}$ are the free parameters,  
$C_1 = 5\times10^{-14}\,h^{-2}\,M_\odot^{-1}\,\mathrm{Mpc}^3$, 
$\rho_c$ is the present critical density, $D(\chi)$ is the linear growth factor, 
and $z_0 = 0.6$ is the pivot redshift.  

\begin{figure*}
    \centering
    \subfigure{\includegraphics[width=0.45\linewidth]{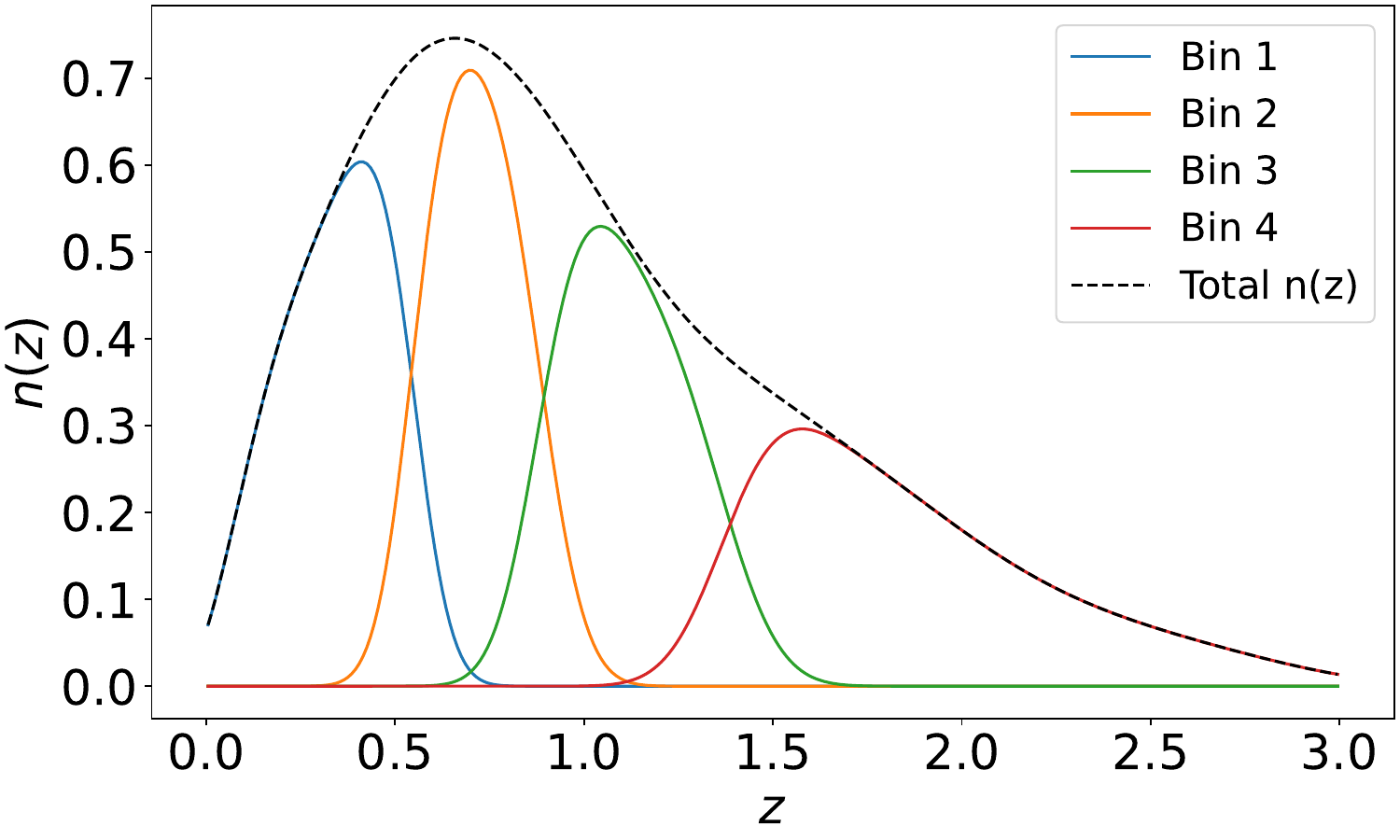}}
    \subfigure{\includegraphics[width=0.45\linewidth]{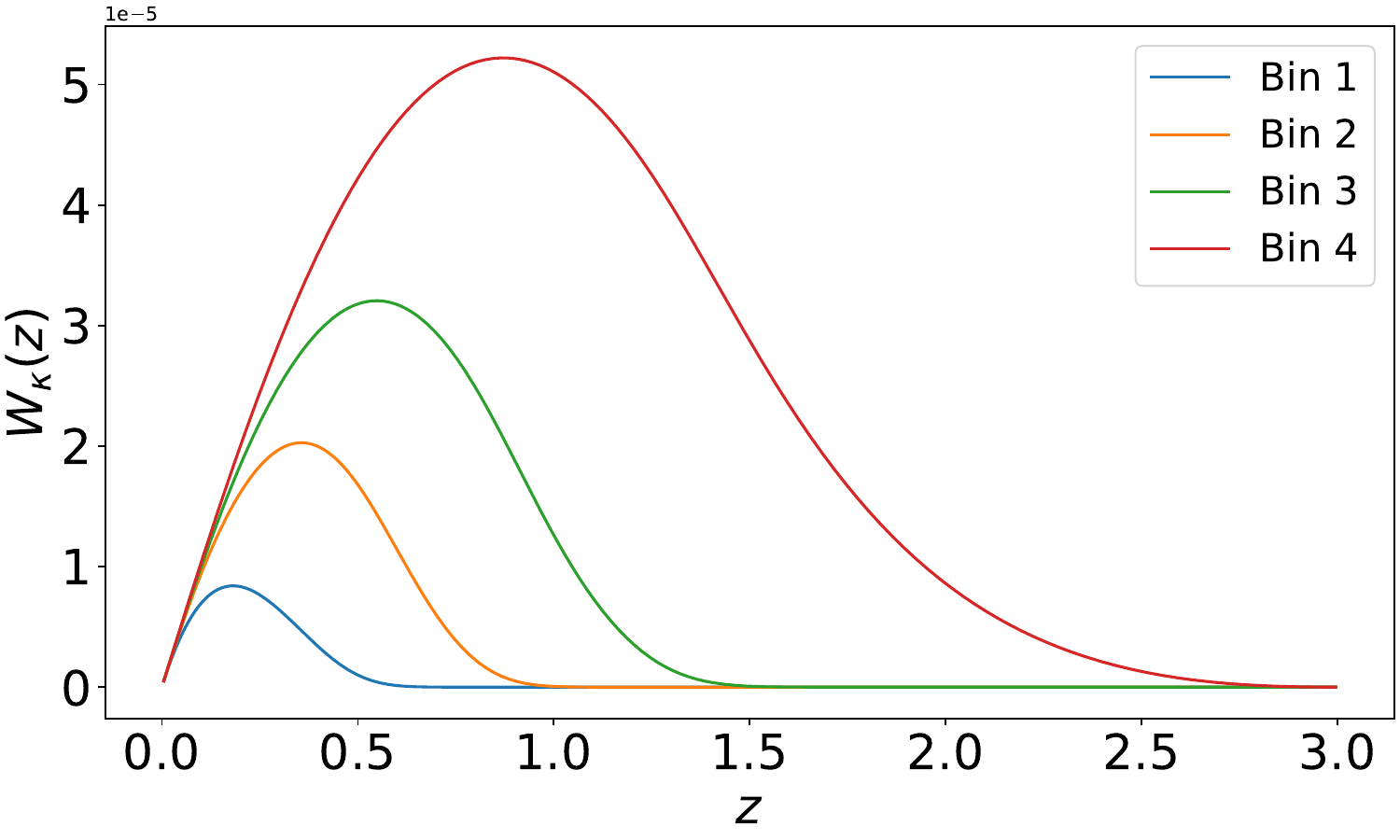}}
    \caption{{\it Left panel}: Galaxy redshift distributions $n(z)$ for the four tomographic bins (colored curves), with the normalized total distribution shown as the black dashed line. {\it Right panel}: The corresponding lensing weight functions $W_\kappa(z)$ for each bin used in the cosmic shear analysis.}
   \label{fig:nzqz}
\end{figure*}

Once all relevant effects are modeled,  
the theoretical angular power spectra are computed  
under the Limber approximation \citep{Limber1953,LoVerde_2008},  
which provides an accurate description on small and  
intermediate angular scales.  
For a generic field $X$ and $Y$,  
the cross spectrum between tomographic bins $i$ and $j$ is
\begin{equation}
C^{ij}_{XY}(\ell) =
\int_0^{\chi_{\text{max}}} \frac{d\chi}{\chi^2} \,
W_X^i(\chi) \, W_Y^j(\chi) \,
P_{\rm m}\!\left( \frac{\ell+1/2}{\chi}, z(\chi) \right),
\end{equation}
where $W_X^i(\chi)$ and $W_Y^j(\chi)$  
are the weighting functions of the two probes.  
The weighting functions for shear ($\kappa$),  
galaxy clustering ($g$), and IA are
\begin{align}
W_\kappa^i(\chi) &=
\frac{3H_0^2\Omega_{\rm m}}{2c^2} \frac{\chi}{a(\chi)}
\int_\chi^{\chi_{\text{max}}} d\chi' \,
n^i(z(\chi')) \frac{dz}{d\chi'}
\frac{\chi' - \chi}{\chi'}, \\
W_g^i(\chi) &= b^i \, n^i(z(\chi)) \frac{dz}{d\chi}, \\
W_{\text{IA}}^i(\chi) &= F^i(\chi).
\end{align}

The full $3\times2$pt vector combines the 
auto- and cross-correlations of galaxy clustering ($gg$),  
cosmic shear ($\kappa\kappa$),  
and galaxy-galaxy lensing ($g\kappa$).  
The observed angular power spectra are given by \citep{Huterer:2005ez,PhysRevD.98.043526}

\begin{align}
\tilde{C}_{gg}^{ij}(\ell) &=
C_{gg}^{ij}(\ell) + \delta_{ij} \frac{1}{\bar{n}^i_g}, \\
\tilde{C}_{\kappa\kappa}^{ij}(\ell) &=
(1+m_i)(1+m_j)
\left[ C_{\kappa\kappa}^{ij}(\ell)
+ C_{\text{II}}^{ij}(\ell)
+ C_{\text{GI}}^{ij}(\ell) \right]  \nonumber \\
&+ \delta_{ij} \frac{\sigma_\gamma^2}{\bar{n}^i_g}, \\
\tilde{C}_{g\kappa}^{ai}(\ell) &=
(1+m_i) \left[
C_{g\kappa}^{ai}(\ell) + C_{g\text{I}}^{ai}(\ell) \right].
\end{align}

Here, $\delta_{ij}$ is the Kronecker delta,
which ensures shot noise and shape noise contributes
only to auto-correlations within
the same tomographic bin, 
$\bar{n}^i_g$ denotes the mean galaxy number density
in the $i$-th bin, and $\sigma_\gamma$ represents the
per-component intrinsic ellipticity dispersion
of source galaxies, characterizing shape noise
in weak-lensing measurements.
These noise and calibration terms are included
to ensure unbiased recovery of
the cosmological power spectra.

\section{Mock Data}
\label{sec:data}

To assess the constraining power of CSST on Hu-Sawicki $f(R)$ gravity, we construct high-fidelity mock catalogs and maps tailored to the CSST photometric survey using the Jiutian simulations \citep{Han:2025fgd,Xiong2025}. These mocks are derived from a large-volume $N$-body simulation with $6144^3$ dark matter particles in a  $1\,h^{-1}\,\mathrm{Gpc}$ box, adopting a fiducial cosmology consistent with Planck 2018 results \citep{Planck2018}. With a particle mass of approximately $3.72\times10^{8} \, h^{-1}\,M_{\odot}$, the simulation captures the nonlinear structure formation essential for our analysis. The mock footprint covers approximately $100\,\mathrm{deg}^2$, matching the scope of the present study.

In the simulation, dark matter evolves within a cubic volume of
order $({\rm Gpc}\,h^{-1})^{3}$. Halos and subhalos are identified
with Friends-of-Friends and SUBFIND, tracing the cosmic web across
time \citep{Springel:2000yr, Springel:2005mi}. 
A semi-analytic galaxy population model, calibrated to deep
multi-band observations, populates halos to match the CSST depth and
bandpasses \citep{Henriques:2014sga, pei2024simulatingemissionlinegalaxies}. 
Each galaxy is assigned a photometric redshift with a scatter of $\sigma_z=0.05(1+z)$. The sample is divided into four tomographic bins spanning $z\in[0,3]$ to extract more information. The resulting effective source density is $\bar{n}_{\mathrm{eff}} \simeq 26\,\mathrm{arcmin}^{-2}$.
The resulting galaxy redshift distribution and weighting functions for 
the galaxy clustering and weak lensing kernels are shown in Figure~\ref{fig:nzqz}.

The weak lensing maps are produced with multi-plane ray tracing along the
light-cone, yielding shear and convergence fields matched to the matter
distribution. Galaxy clustering maps are obtained by projecting the
three-dimensional overdensity in each tomographic slice onto the sky \citep{Chen:2023jiy}. 
These products yield realistic $\kappa\kappa$, $gg$, and
$g\kappa$ signals suited to a full $3\times2$pt analysis.

The maps are rendered on uniform Cartesian grids with angular resolution
sufficient for $\ell_{\max}\!\approx\!3000$. 
The noise realizations follow the CSST characteristics and are derived 
based on the results from current observations. We adopt the 
per-component variance $\sigma_{\gamma}^{2}=0.04$ for the shape noise 
in the shear maps and the Poisson 
shot noise is added to the galaxy clustering maps in each redshift bin.
Our noise realizations are uncorrelated between tomographic bins, and shot
noise contributes only to auto-correlations.

We estimate the angular power spectra $C_{\ell}$ for all auto- and
cross-correlations on flat-sky patches using Fourier estimators, with
logarithmically spaced multipole bins over $\ell\in[50,3000]$ to balance
large-scale sensitivity and small-scale modeling. Mode coupling from
masks is handled through pseudo-$C_\ell$ deconvolution \citep{annurev:/content/journals/10.1146/annurev-astro-081811-125526, Hivon_2002,10.1111/j.1365-2966.2004.07737.x,Hikage_2010}.

\begin{figure}
    \centering
    \includegraphics[width=\linewidth]
    {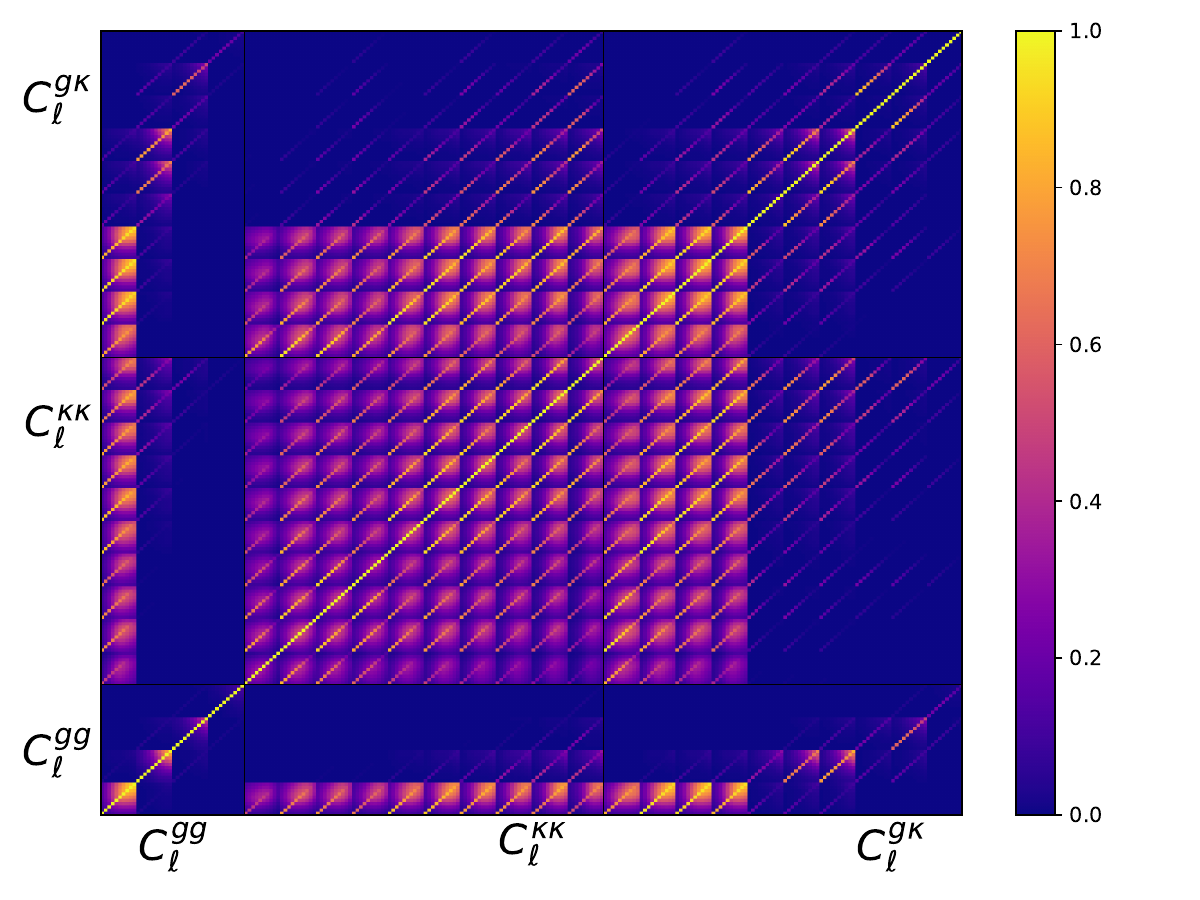}
    \caption{The normalized covariance matrix for the joint $3\times2$pt data vector, structured by tomographic bin and probe type ($gg$, $\kappa\kappa$, $g\kappa$). The matrix shows the correlation coefficients between all measured angular power spectrum combinations.}
    \label{fig:covmat}
\end{figure}

The joint covariance of the $3\times2$pt data vector
is decomposed into three components: Gaussian,
connected non-Gaussian (cNG) 
\citep{Hamana:2003ts,Takada:2003ef,Huterer:2005ez}, and super-sample
covariance (SSC) \citep{Hu:2003pt,Takada:2013wfa}. 
Following the methodology validated in \citet{Xiong2025}, we have
$
\mathrm{Cov}
= \mathrm{Cov}^{\mathrm{Gauss}}
+ \mathrm{Cov}^{\mathrm{cNG}}
+ \mathrm{Cov}^{\mathrm{SSC}}.
$
The Gaussian term is computed analytically as the
four-point contraction of two power spectra, which is given by
\begin{equation}
\begin{split}
&\mathrm{Cov}^{\rm Gauss}_{ij,kl;XY,X'Y'}(\ell,\ell')
= \frac{\delta^{\mathrm{K}}_{\ell\ell'}}
{(2\ell+1) f_{\mathrm{sky}} \Delta\ell} \\
&\times
\Big[
\tilde{C}^{ik}_{XX'}(\ell)\,
\tilde{C}^{jl}_{YY'}(\ell) 
+ \tilde{C}^{il}_{XY'}(\ell)\,
\tilde{C}^{jk}_{YX'}(\ell)
\Big],
\end{split}
\label{eq:cov_gauss}
\end{equation}
where $X, X', Y, Y' \in \{g, \kappa\}$ denote the
tracer types (galaxy or convergence),
$i, j, k, l$ index the tomographic bins,
$f_{\mathrm{sky}}$ is the survey sky fraction,
and $\Delta\ell$ is the multipole bin width.
For non-Gaussian terms,
the cNG term captures mode coupling arising
from nonlinear structure formation and becomes
increasingly important at high multipoles, and  
the SSC term quantifies the coupling between modes within the survey and
long-wavelength background density fluctuations.
The non-Gaussian covariance components are computed
using the CCL code \citep{Chisari_2019,CCL_GitHub}, and
the normalized covariance matrix is shown in
Figure~\ref{fig:covmat}.

\begin{figure*}
    \centering
    \includegraphics[width=0.85\linewidth]{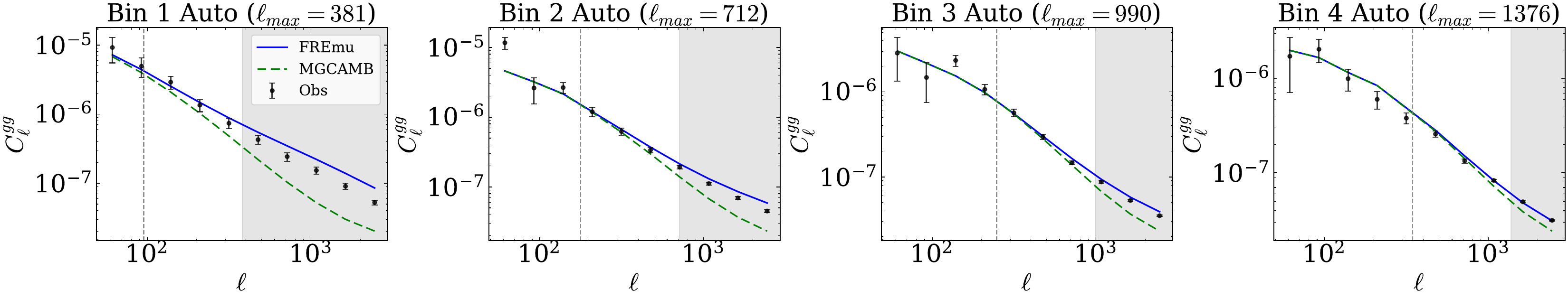}
    \caption{The simulated angular power spectra $C_{gg}^{ij}(\ell)$ of the CSST galaxy clustering
     survey for the four tomographic bins. The solid blue and dashed green curves show the prediction
     using \textsc{FREmu} and \textsc{MGCAMB} at $\log_{10}|f_{R0}|=-6$, respectively. The error bars
     are derived from the diagonal of the covariance matrix. 
     The gray region indicates the multipoles excluded by the scale cut at 
     $k=0.4 \, h \, \mathrm{Mpc}^{-1}$. The vertical gray dashed
     lines mark $\ell_{\max}$ corresponding to $k = 0.1\,h\,\mathrm{Mpc}^{-1}$. }
    \label{fig:clgg}
\end{figure*}

\begin{figure*}
    \centering
    \includegraphics[width=0.85\linewidth]{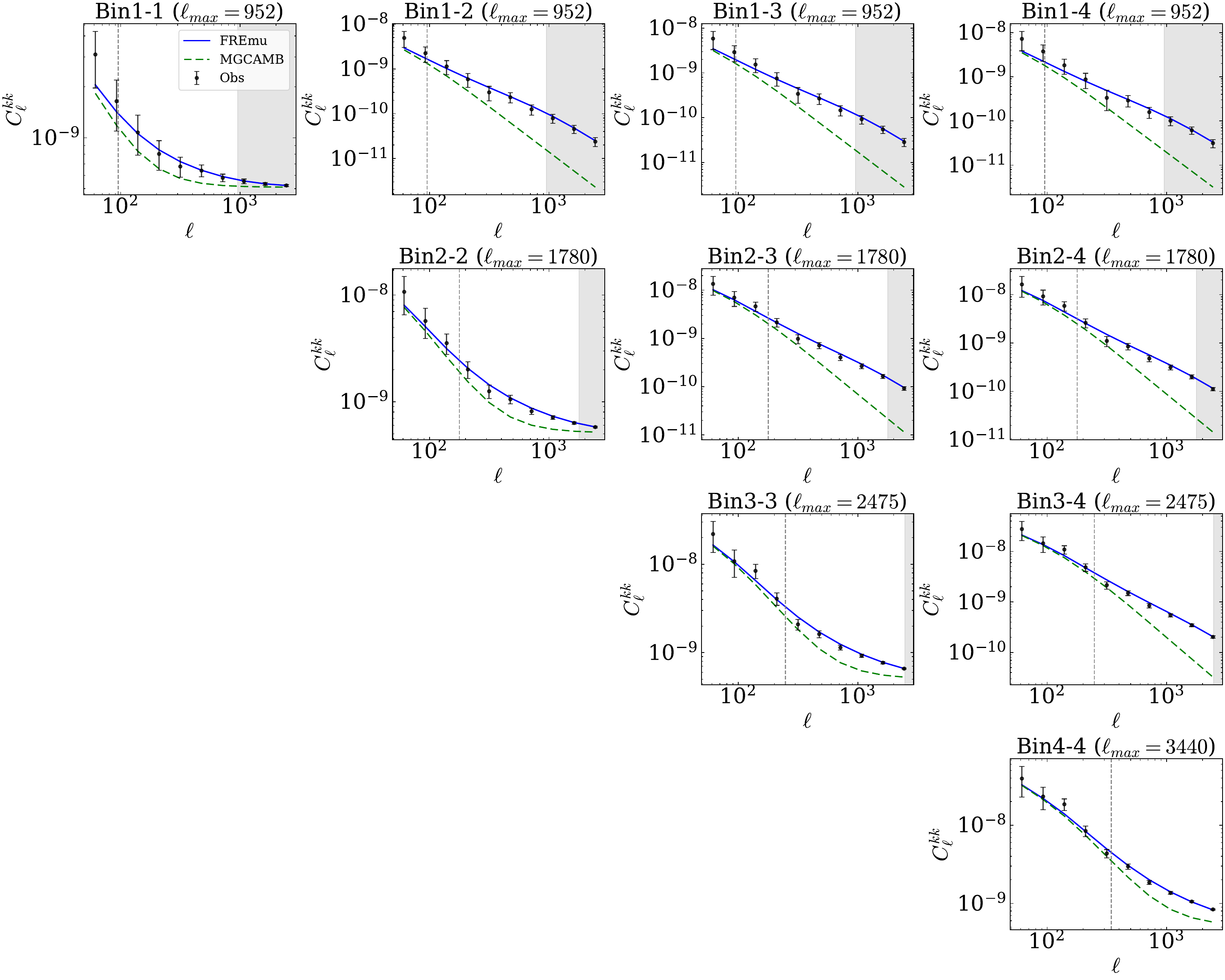}
    \caption{The simulated angular power spectra $C_{\kappa\kappa}^{ij}(\ell)$ of the CSST shear  
    measurement for the four tomographic bins. The solid blue and dashed green curves show the prediction
     using \textsc{FREmu} and \textsc{MGCAMB} at $\log_{10}|f_{R0}|=-6$, respectively. The error bars
     are derived from the diagonal of the covariance matrix.  The gray region indicates the multipoles 
     excluded by the scale cut at 
     $k=1.0 \, h \, \mathrm{Mpc}^{-1}$.. The vertical gray dashed
     lines mark $\ell_{\max}$ corresponding to $k = 0.1\,h\,\mathrm{Mpc}^{-1}$. }
    \label{fig:clkk}
\end{figure*}

\begin{figure*}
    \centering
    \includegraphics[width=0.85\linewidth]{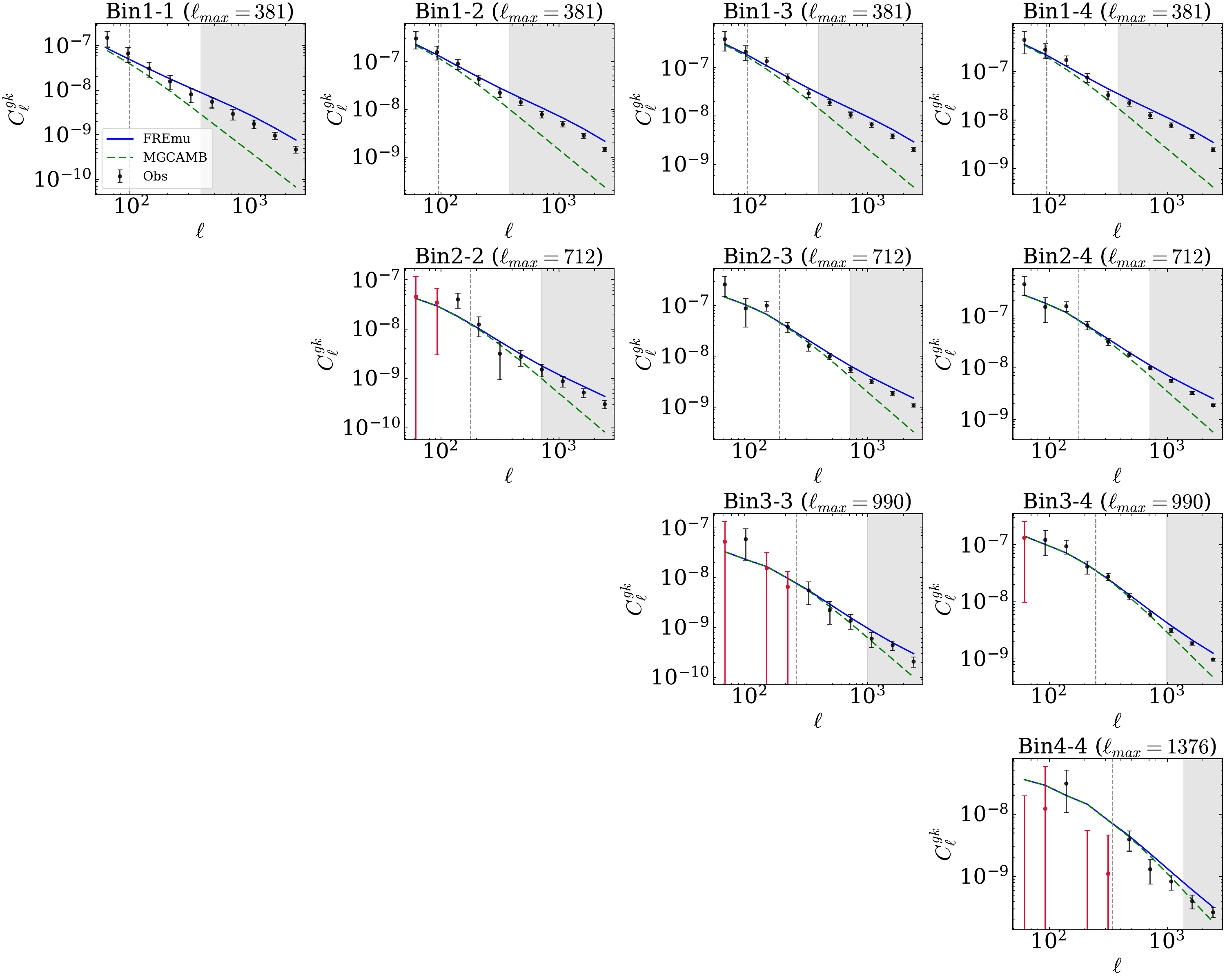}
    \caption{The simulated angular power spectra $C_{g\kappa}^{ij}(\ell)$ of the CSST galaxy-galaxy lensing   
    measurement for the four tomographic bins. The solid blue and dashed green curves show the prediction
     using \textsc{FREmu} and \textsc{MGCAMB} at $\log_{10}|f_{R0}|=-6$, respectively. The error bars
     are derived from the diagonal of the covariance matrix.  
     The gray region indicates the multipoles excluded by the scale cut at 
     $k=0.4 \, h \, \mathrm{Mpc}^{-1}$. The vertical gray dashed
     lines mark $\ell_{\max}$ corresponding to $k = 0.1\,h\,\mathrm{Mpc}^{-1}$. Data points with signal-to-noise ratio (SNR) below unity are highlighted in red to indicate low statistical significance.}
    \label{fig:clgk}
\end{figure*}

In Figures~\ref{fig:clgg}, \ref{fig:clkk}, and \ref{fig:clgk}, we show the
observed galaxy, shear, and galaxy-galaxy lensing angular power spectra, respectively. 
The $1\sigma$ error bars are from the diagonal of the covariance matrix. The theoretical curves from two parallel pipelines are also shown, i.e. a linear theory pipeline using \textsc{MGCAMB} 
\citep{Lewis_2000,Zhao_2009,Zucca_2019,Hojjati_2011,Wang_2023}, and a 
nonlinear pipeline using the \textsc{FREmu} emulator \citep{Bai2024,Bai:2024hpw}.
To control modeling systematics, we impose $k$-space cuts and map
them to per-bin multipole cuts via
$\ell_{\max}^{(i)}=k_{\max}\,\chi(\langle z^{(i)}\rangle)$. In the linear pipeline, we adopt $k_{\max}=0.1,h,\mathrm{Mpc}^{-1}$ for all probes, which corresponds to the vertical gray dashed lines in Figures~\ref{fig:clgg}, \ref{fig:clkk}, and \ref{fig:clgk}. In the emulator pipeline, we use
$k_{\max}=0.4\,h\,\mathrm{Mpc}^{-1}$ for $gg$ and $g\kappa$ to limit
the impact of small scale uncertainties such as nonlinear screening
residuals and potential baryonic effects, and set
$k_{\max}=1.0\,h\,\mathrm{Mpc}^{-1}$ for $\kappa\kappa$, which is determined by
the \textsc{FREmu} training upper bound. The gray shaded regions in 
Figures~\ref{fig:clgg}, \ref{fig:clkk}, and \ref{fig:clgk} indicate the 
multipoles excluded by the applicable cuts.

\section{Parameter Constraints}
\label{sec:constraints}
\subsection{Constraint Process}

We employ a Bayesian framework to infer the posterior 
distributions of the cosmological, modified gravity, 
and systematic parameters from the measured angular 
power spectra, using Markov chain Monte Carlo (MCMC) 
sampling. To structure the analysis, we define 
data vectors 
$\vec{D}_X = {\tilde{C}^{ij}_X(\ell)}$, 
where $X \in {gg, \kappa\kappa, g\kappa}$, which 
concatenate all tomographic bin pairs and multipoles 
for each probe. The log-likelihood for each probe 
is assumed to be Gaussian and is given by
\begin{equation}
\ln\mathcal{L}_X
= -\tfrac{1}{2}\big(\vec{D}_X - \vec{T}_X(\boldsymbol{\theta})\big)^{\mathrm{T}}
\mathbf{C}_X^{-1}\big(\vec{D}_X - \vec{T}_X(\boldsymbol{\theta})\big),
\end{equation}
where $\vec T_X(\boldsymbol{\theta})$ are the theoretical angular power
spectra considering the systematic terms, and $\mathbf C_X$ are the corresponding
covariances including the Gaussian and non-Gaussian terms (cNG and SSC),
with the Gaussian part given by Eq.~(\ref{eq:cov_gauss}).

The posterior follows Bayes' theorem with priors specified in
Table~\ref{tab:free_parameters}.
The priors of the cosmological parameters and modified gravity amplitude we adopt
are fully contained within the \textsc{FREmu} training
hypercube, ensuring emulator validity across the sampled parameter
space. We vary the intrinsic
alignment amplitude $A_{\rm IA}$ and its redshift evolution $\eta_{\rm IA}$, 
acknowledging that the present setup
has limited constraining power on $\eta_{\rm IA}$. For this mock-based
study we disable explicit modeling of baryonic feedback, since the mocks
do not include baryonic physics and our scale selection reduces baryonic
sensitivity.

\begin{table}[h]
\centering
\caption{The free parameters considered in our constraint process. 
The first column shows the names of the free parameters, and the 
second and third columns show the fiducial values and the priors 
of the parameters, respectively. Uniform priors are described by 
$\mathcal{U}(x, y)$, with $x$ and $y$ denoting the prior range. 
The Gaussian priors are represented by $\mathcal{N}(\mu, \sigma)$, 
where $\mu$ and $\sigma$ are the mean and standard deviation, 
respectively.}
\begin{tabular}{lll}
\hline
\hline
Parameter & Fiducial Value & Prior \\
\hline
\multicolumn{3}{l}{\textbf{Cosmology}} \\

$H_0$ & 67.66 & $\mathcal{U}(50, 90)$ \\

$\Omega_{\rm m}$ & 0.3111 & $\mathcal{U}(0.1, 0.5)$ \\

$\Omega_{\rm b}$ & 0.0490 & $\mathcal{U}(0.03, 0.07)$ \\

$\log(10^{10} A_{\rm s})$ & 3.05 & $\mathcal{U}(1.61, 3.91)$ \\

$n_{\rm s}$ & 0.9665 & $\mathcal{U}(0.8, 1.2)$ \\

$\log_{10} |f_{R0}|$ & - & $\mathcal{U}(-7.0, -3.5)$ \\
\hline
\multicolumn{3}{l}{\textbf{Intrinsic alignment}} \\

$A_{\rm IA}$ & 0 & $\mathcal{U}(-5, 5)$ \\

$\eta_{\rm IA}$ & 0 & $\mathcal{U}(-5, 5)$ \\
\hline
\multicolumn{3}{l}{\textbf{Galaxy bias}} \\

$b_g^i$ & - & $\mathcal{U}(0, 5)$ \\
\hline
\multicolumn{3}{l}{\textbf{Photo-z shift}} \\

$\Delta z^i$ & (0, 0, 0, 0) & $\mathcal{N}(0, 0.01)$ \\
\hline
\multicolumn{3}{l}{\textbf{Photo-z stretch}} \\

$\sigma_z^i$ & (1, 1, 1, 1) & $\mathcal{N}(1, 0.05)$ \\

\multicolumn{3}{l}{\textbf{Shear calibration}} \\
\hline
$m_i$ & (0, 0, 0, 0) & $\mathcal{N}(0, 0.01)$ \\
\hline
\hline
\end{tabular}
\label{tab:free_parameters}
\end{table}

As mentioned, we consider two routes to compute the theoretical angular power spectra
that enter the likelihood. The first route uses \textsc{MGCAMB} , which solves
the linearized Boltzmann-Einstein equations in the Hu-Sawicki model,
yields the linear matter power spectrum, and provides modified growth and
lensing responses at the perturbative level. We then project the resulting
spectra to obtain $C_\ell$ for $gg$, $\kappa\kappa$, and $g\kappa$ under
the same tomographic windows used in the measurements. Since a validated
nonlinear prescription for modified gravity is not included in this setup,
we restrict to $k_{\max}=0.1\,h\,\mathrm{Mpc}^{-1}$ for all probes, 
as indicated by the gray dashed lines in Figures~\ref{fig:clgg}-\ref{fig:clgk}. This route serves as a conservative reference that isolates information from the linear regime.

The second route employs the emulator \textsc{FREmu}, which returns the ratio
$B(k,z)\equiv P_{f(R)}(k,z)/P_{\Lambda\mathrm{CDM}}(k,z)$ over a validated
training domain in parameter space, redshift, and wavenumber. We obtain a
nonlinear prediction for $f(R)$ by multiplying $B(k,z)$ with a baseline
nonlinear prescription for $\Lambda$CDM. This construction preserves the
large scale agreement with the linear route and extends to smaller
scales under controlled modeling. We have verified that the results of the two routes 
are in good agreement on the overlapping large-scale range, as shown in
Figures~\ref{fig:clgg}-\ref{fig:clgk}, while the emulator can correctly retain 
additional high $\ell$ information.

For parameter inference, we use \textsc{Cobaya} 
\citep{Torrado_2021,Cobaya_ASCL_1910019} to run MCMC sampling. We
implement probe specific likelihoods for $gg$, $\kappa\kappa$, and
$g\kappa$ under both routes. We assess the convergence of 
our Markov chains using the Gelman-Rubin statistic quantity $R$. Chains are 
considered converged when $R-1 < 0.01$. To minimize the influence 
of the initial sampling phase, we discard the first $30\%$ of each 
chain as burn-in and use the remaining samples for posterior inference.

\subsection{Constraint Results}
\begin{figure}
    \centering
    \includegraphics[width=\linewidth]{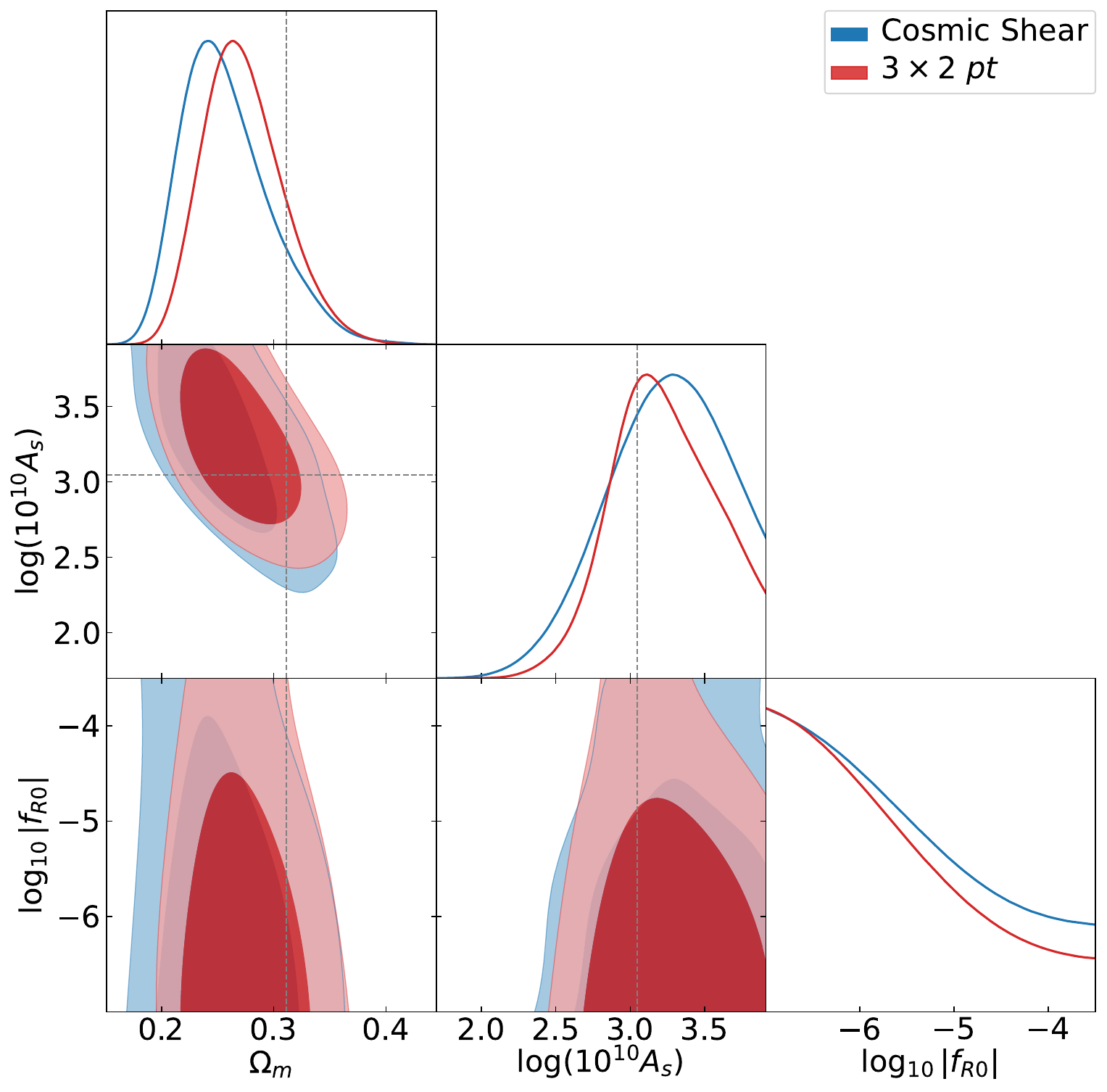}
    \caption{The contour maps (68\% and 95\% CLs) and 1D PDFs of $\Omega_{\mathrm m}$, $\log(10^{10} A_{\rm s})$, $\log_{10}|f_{R0}|)$ for the CSST shear and 3$\times$2pt measurements in 100 deg$^2$ from the \textsc{FREmu} emulator. The dashed lines indicate the fiducial values of the parameters.}
    \label{fig:contour_map_no_H0_ns}
\end{figure}

Our analysis reveals a stark contrast in constraining power between the two theoretical approaches. The \textsc{MGCAMB}-based linear analysis fails to provide a meaningful constraint on $\log_{10}|f_{R0}|$, as the Hu-Sawicki model's deviations from GR are subtle on the linear scales ($k_{\text{max}} = 0.1\, h\,\mathrm{Mpc}^{-1}$) accessible to this method and are dwarfed by statistical uncertainties. In contrast, the \textsc{FREmu} emulator, by incorporating nonlinear information, enables a significant detection of the modified gravity signal.

The resulting constraints from the \textsc{FREmu}-based analysis are summarized in Figure~\ref{fig:contour_map_no_H0_ns}. For the cosmic shear-only ($\kappa\kappa$) case, we find $\log_{10}|f_{R0}| < -5.29$ ($-3.83$) at the 68\% (95\%) confidence level. The joint $3\times2$pt analysis tightens this constraint to $\log_{10}|f_{R0}| < -5.42$ ($-3.88$). Beyond the modified gravity parameter, the joint analysis also improves the precision on the standard $\Lambda$CDM cosmology: the relative uncertainty on $\Omega_{\mathrm{m}}$ improves from $15.4\%$ to $13.0\%$, and on $\log(10^{10}A_{\rm s})$ from $11.8\%$ to $10.3\%$.

It is worth noting that the 100 deg$^2$ dataset provides relatively weak constraints on other cosmological parameters such as $H_0$, $\Omega_\mathrm{b}$, and $n_s$, as shown in the full posterior distributions in the Appendix. This is reasonable given the limited sky coverage and the inherent degeneracies in the current $3\times2$pt data vector.

The constraining power is expected to increase dramatically with the full CSST photometric survey covering 17,500 deg$^2$. The enhanced measurement precision on large-scale modes will be particularly beneficial for galaxy clustering, and the constraint on $\log_{10}|f_{R0}|$ could be improved by approximately an order of magnitude in this case. Furthermore, the full survey will also provide significantly improved constraints on parameters like $H_0$, $\Omega_\mathrm{b}$, and $n_s$ through better measurement of the large-scale angular power spectra. This positions the full CSST survey to deliver exceptionally stringent tests of the Hu-Sawicki model and to break degeneracies in the full cosmological parameter space.

These results demonstrate that the CSST $3\times2$pt analysis can achieve competitive constraints using its internal data alone. This capability is noteworthy, as current Stage-III surveys often find that cosmic shear alone yields prior-dominated constraints on $f(R)$ gravity, frequently requiring combination with external datasets like CMB and BAO to achieve stringent bounds \citep{Vazsonyi2021, Bai:2024hpw, SPT:2024adw}.

\section{Summary}
\label{sec:summary}

In this work, we have established a robust analysis framework to assess the potential of the CSST $3\times2$pt photometric survey for constraining the Hu Sawicki $f(R)$ gravity model. Our approach integrates high fidelity mock observations that capture the essential features of the survey, including photometric redshift sampling, shape and shot noise, and mask induced mode coupling. We measured the angular power spectra for galaxy clustering, cosmic shear, and galaxy galaxy lensing from these mocks. For theoretical predictions, we employed two complementary methods: a linear theory approach using \textsc{MGCAMB} restricted to conservative scales of $k_{\max}=0.1,h,\mathrm{Mpc}^{-1}$, and a nonlinear approach utilizing the \textsc{FREmu} emulator, which extends to $k_{\max}=0.4,h,\mathrm{Mpc}^{-1}$ for galaxy clustering and galaxy galaxy lensing, and $k_{\max}=1.0,h,\mathrm{Mpc}^{-1}$ for cosmic shear. The two methods show good agreement on overlapping large scales, with the emulator successfully incorporating additional information from smaller scales.

We performed parameter inference with the MCMC sampler in \textsc{Cobaya}, using probe specific likelihoods and a block diagonal covariance approximation. The cosmological and modified gravity parameters were sampled within the validated domain of the \textsc{FREmu} emulator, while simultaneously marginalizing over a comprehensive set of systematic parameters, including those for intrinsic alignment, photometric redshift uncertainties, galaxy bias, and shear calibration. Our analysis shows that the CSST $3\times2$pt 100 deg$^{2}$ survey delivers constraints on $\log_{10}|f_{R0}|$ that are competitive with, and in fact slightly tighter than, those from current Stage III surveys.

The constraining power is projected to improve substantially with the full CSST photometric survey covering 17,500 deg$^2$. Future analyses combining this dataset with external probes such as CMB lensing and cluster abundances are expected to enable exceptionally stringent tests of modified gravity models and possible deviations from General Relativity on cosmological scales.


\begin{acknowledgments}
J.H.Y and Y.G. acknowledge the support from the CAS Project for Young Scientists in Basic Research (No. YSBR-092), and National Key R\&D Program of China grant Nos. 2022YFF0503404 and 2020SKA0110402. X.L.C. acknowledges the support of the National Natural Science Foundation of China through grant Nos. 11473044 and 11973047 and the Chinese Academy of Science grants ZDKYYQ20200008, QYZDJ- SSW-SLH017, XDB 23040100, and XDA15020200. Q.G. acknowledges the support from the National Natural Science Foundation of China (NSFC No. 12033008). The Jiutian simulations were conducted under the support of the science research grants from the China Manned Space Project with grant No. CMS- CSST-2021-A03. This work is also supported by science research grants from the China Manned Space Project with grant Nos. CMS-CSST-2025-A02, CMS-CSST-2021-B01, and CMS-CSST-2021-A01.
\end{acknowledgments}

%

\software{
    \textsc{Cobaya} \citep{Torrado_2021,Cobaya_ASCL_1910019}, 
\textsc{MGCAMB} \citep{Lewis_2000,Zhao_2009,Zucca_2019,Hojjati_2011,Wang_2023}, 
\textsc{CCL} \citep{Chisari_2019,CCL_GitHub}, 
\textsc{FREmu}\citep{Bai2024,Bai:2024hpw},
\textsc{getdist}\citep{Lewis:2019xzd}
          }


\appendix

\begin{figure}
    \centering
    \includegraphics[width=\linewidth]{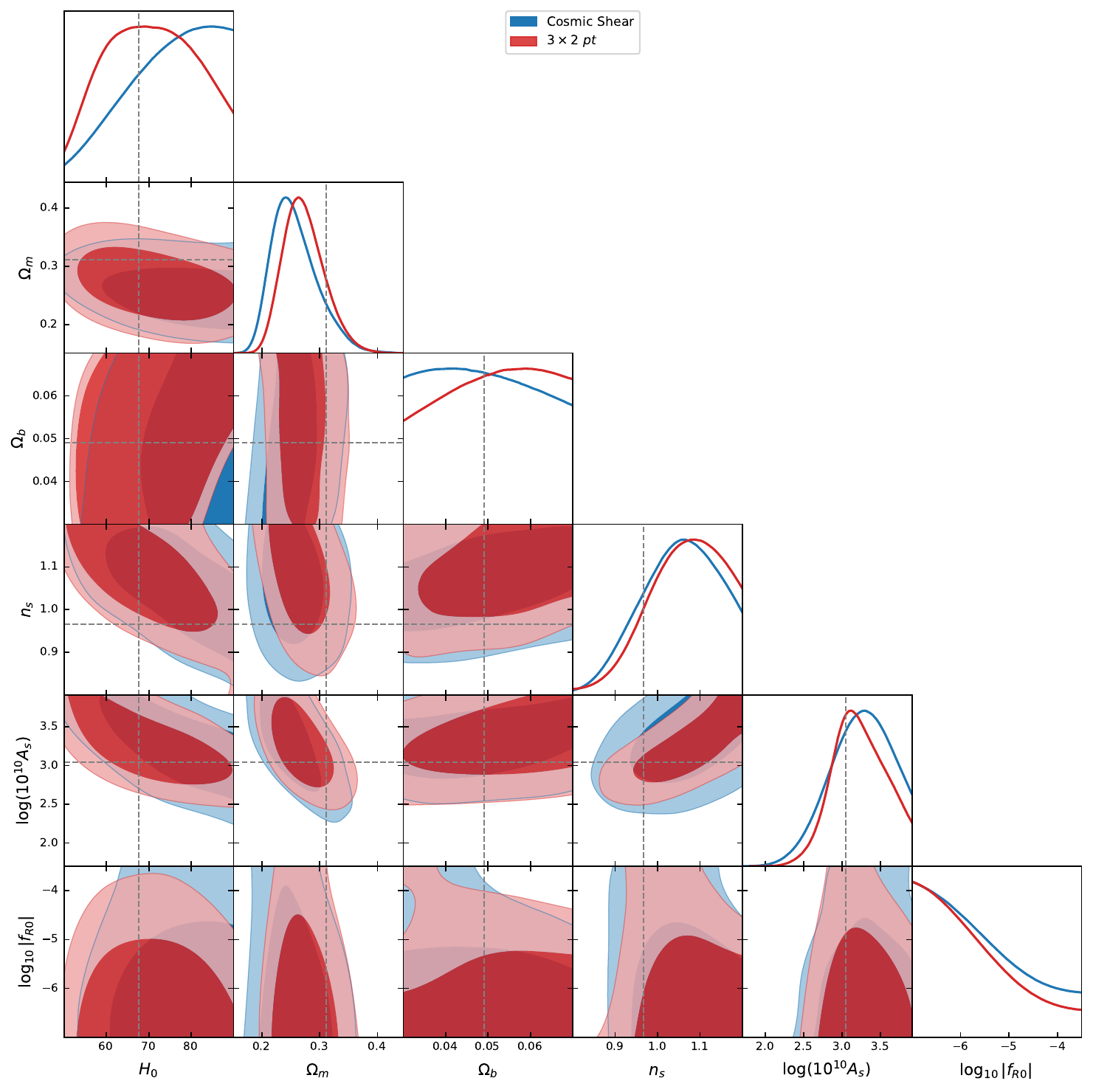}
    \caption{The contour maps (68\% and 95\% CLs) and 1D PDFs of all cosmological parameters considered in this work for the CSST shear and 3$\times$2pt measurements in 100 deg$^2$ from the \textsc{FREmu} emulator. The dashed lines indicate the fiducial values of the parameters.}
    \label{fig:contour_map_all_cosmo_paras}
\end{figure}

\begin{figure}
    \centering
    \includegraphics[width=\linewidth]{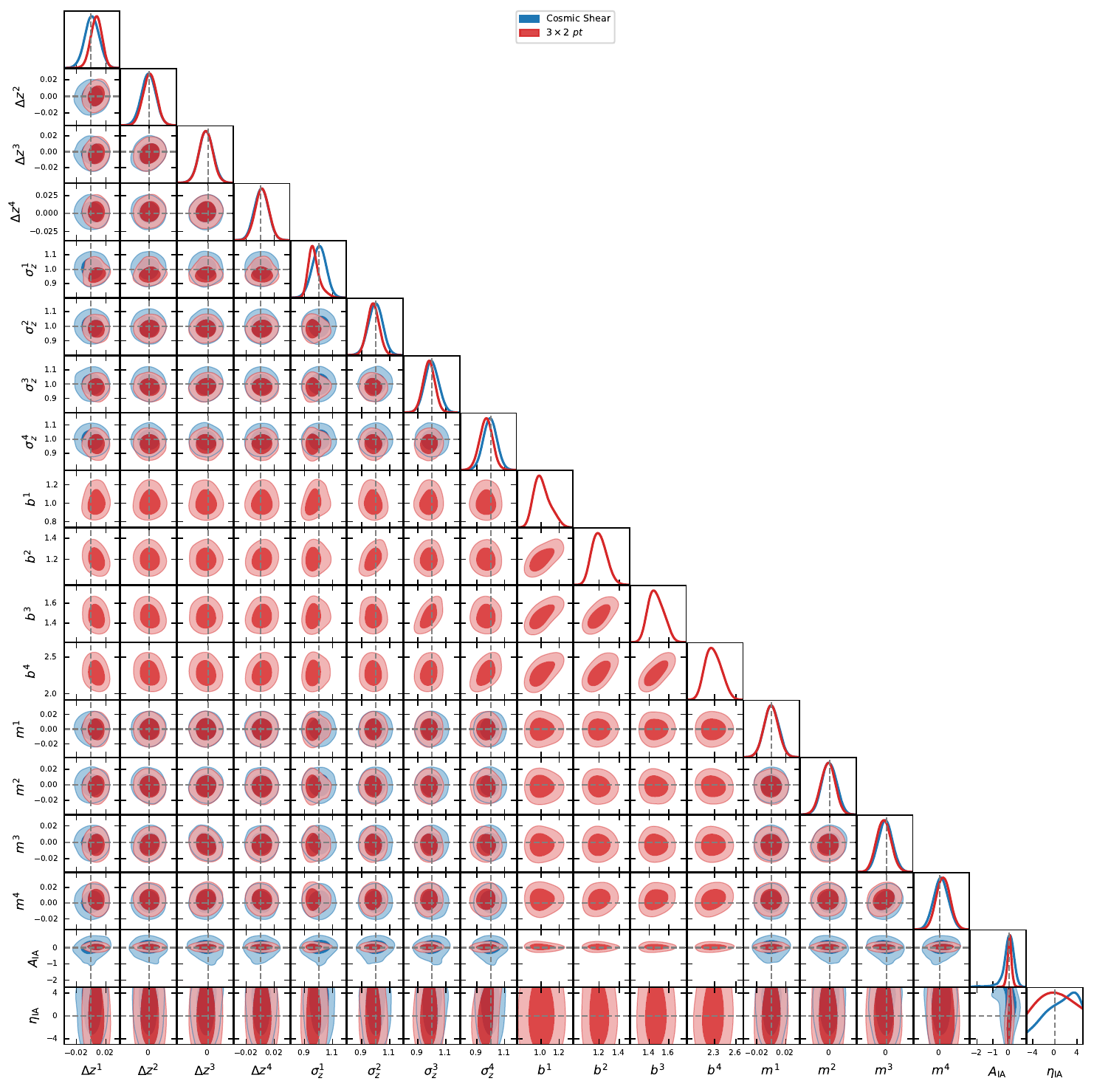}
    \caption{The contour maps (68\% and 95\% CLs) and 1D PDFs of the systematic parameters for the intrinsic alignment ($A_{\rm IA}$, $\eta_{\rm IA}$), photo-$z$ shift $\Delta z^i$ and stretch $\sigma_z^i$, galaxy bias $b_g^i$, and shear calibration $m_i$ for the CSST shear and 3$\times$2pt measurements in 100 deg$^2$ from the \textsc{FREmu} emulator. The dashed lines indicate the fiducial values of the parameters.}
    \label{fig:contour_map_nuisance_paras}
\end{figure}

In this appendix we present the full posterior constraints for 
all cosmological parameters (Figure~\ref{fig:contour_map_all_cosmo_paras}) and systematic parameters (Figure~\ref{fig:contour_map_nuisance_paras}), under the \textsc{FREmu} emulator method. Unless 
otherwise stated, the chains adopt the scale selection described 
in the main text (with $gg$ and $g\kappa$ restricted to 
$k_{\max}=0.4\,h\,\mathrm{Mpc}^{-1}$ and $\kappa\kappa$ to 
$k_{\max}=1.0\,h\,\mathrm{Mpc}^{-1}$). We show the constraint results 
for the cosmic shear only ($\kappa \kappa$) and the joint $3\times2$pt combination 
($gg, \ g\kappa, \ \kappa\kappa$) cases. 
These figures complement the focused three-parameter result shown in 
Figure~\ref{fig:contour_map_no_H0_ns}, and provide a complete picture 
of systematic and cosmological degeneracies under the adopted likelihood 
and covariance matrix.

\bibliography{sample701}{}
\bibliographystyle{aasjournalv7}



\end{document}